\documentclass[aps,prb,preprint,showpacs,byrevtex]{revtex4}

\usepackage{times,mathptm}
\usepackage[dvips]{graphicx,color}

\begin{document}

\title{Antiferromagnetic ordering of dangling-bond electrons at the stepped Si(001) surface}
\author{Jun-Ho Lee, Sun-Woo Kim, and Jun-Hyung Cho$^*$}
\affiliation{Department of Physics and Research Institute for Natural Sciences, Hanyang University,
17 Haengdang-Dong, Seongdong-Ku, Seoul 133-791, Korea}
\date{\today}

\begin{abstract}
Using first-principles density-functional calculations, we explore the possibility of magnetic order at the rebonded $D_B$ step of the Si(001) surface. The rebonded $D_B$ step containing threefold coordinated Si atoms can be treated as a one-dimensional dangling-bond (DB) wire along the step edge.
We find that Si atoms composing the step edge are displaced up and down alternatively due to Jahn-Teller-like distortion, but, if Si dimers on the terrace are passivated by H atoms, the antiferromagnetic (AFM) order can be stabilized at the step edge with a suppression of Jahn-Teller-like distortion. We also find that the energy preference of AFM order over Jahn-Teller-like distortion is enhanced in an oscillatory way as the length of DB wires decreases, showing the so-called quantum size effects.
\end{abstract}


\maketitle

\section{INTRODUCTION}

Creation of a magnetically ordered state in carbon materials has attracted much attention because of its fundamental scientific interest as well as for the potential applications in
diverse fields.~\cite{yaz} Indeed, a number of theoretical and experimental studies have been performed to explore the possibility of magnetic orders in various carbon materials such as graphite,~\cite{esq} carbon nanofoams,~\cite{rod} nanodiamonds,~\cite{bar} fullerenes,~\cite{lee2,choi} and graphene.~\cite{son,fuj,nak} For instance, it was theoretically predicted that graphene flakes or nanoribbons with zigzag edges have a high density of states at the Fermi level, therefore yielding a ferromagnetic order at the edge.~\cite{son,fuj,nak}

Since silicon belongs to the same group IV elements with carbon, it is very interesting to extend the discussion of magnetic order in silicon materials. Recently, a first-principles density-functional theory (DFT) calculation of Erwin and Himpsel~\cite{erw} predicted that Au-induced vicinal Si surfaces such as Si(553)-Au and Si(557)-Au exhibit an antiferromagnetic (AFM) coupling between the magnetic moments arising from the Si dangling-bond (DB) electrons. Here, the critical structural motif for the presence of AFM order is a graphitic honeycomb ribbon structure of Si atoms at the step edge. In this sense, the novel way of manipulating magnetism in silicon materials is to create one-dimensional (1D) nanostructures at step edges, which provide more favorable conditions for long-range structural order with an array of spins.

In this paper, we explore the possibility of magnetic orders at the stepped Si(001) surface using first-principles DFT calculations. Here, we consider the double-atomic height step, i.e., the so-called rebonded $D_B$ step [see Fig. 1(a)], which prevails on the Si(001) surface.~\cite{cha} This step contains threefold coordinated Si atoms, giving rise to a 1D array of DBs along the step. Our spin-polarized DFT calculations for such a 1D DB wire show that the AFM or ferromagnetic (FM) configuration, where neighboring spins of DB electrons are coupled antiferromagnetically or ferromagnetically with each other, is energetically unfavored over the nonmagnetic (NM) configuration, where Si atoms composing the rebonded $D_B$ step are alternatively displaced up and down due to Jahn-Teller-like distortion. However, we find that, if Si dimers on the terrace are passivated by H atoms, DB electrons at the step become more localized to yield the preference of the AFM order with a suppression of Jahn-Teller-like distortion. In addition, we find that, as the length of DB wires decreases, the energy preference of the AF configuration over the NM one increases in an oscillatory way. Our findings demonstrate that the AFM spin ordering of DB electrons at the rebonded $D_B$ step can be stabilized by passivating the terrace with atomic H, and its strength can be manipulated with respect to the length of DB wires.

\begin{figure}[hb]
\centering{ \includegraphics[width=7.0cm]{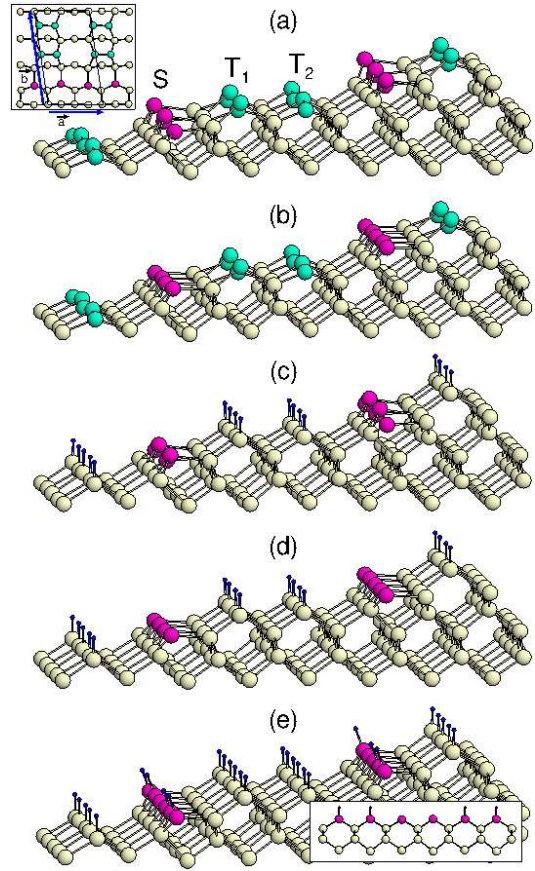} }
\caption{(color online)
Optimized structures for the rebonded $D_B$ step of a Si(117) surface within (a) the NM configuration and
(b) the AFM configuration. The large and small circles represent Si and H atoms, respectively. For distinction, Si atoms at the step edge (denoted by S) and Si dimers on the terrace (denoted by T$_1$ and T$_2$) are drawn in different dark colors.
The optimized structures of an infinitely long DB wire within the NM and AFM
configurations as well as a finitely long DB wire (DB-2) within the AFM configuration
are also given in (c), (d), and (e), respectively. The insets in (a) and (e) show the top view and the side view of each structure, respectively.}
\end{figure}

\section{CALCULATION METHOD}

The total-energy and force calculations were performed by using nonspin-polarized and spin-polarized DFT~\cite{hoh,koh}
within the generalized-gradient approximation.~\cite{per} The Si and H atoms
were described by norm-conserving pseudopotentials.~\cite{tro} To simulate the rebonded
$D_B$ step, we used a triclinic supercell with the lattice vectors of $\vec{a}$ = ($\sqrt{2}$, 0, 0),
$\vec{b}$ = ($-$1/2$\sqrt{2}$, 7/2$\sqrt{2}$, 1/2), and $\vec{c}$ = (0, 0, 17/4) in units of the optimized Si lattice constant $a_0$ = 5.480 {\AA} [see Fig. 1(a)], corresponding to a vicinal surface with Miller indices (117).
The stepped Si(001) surface was modeled by a periodic slab geometry containing
a thickness of six Si atomic layers along the [001] direction.
In this periodic slab geometry, periodically repeated rebonded $D_B$ steps were
separated by terraces with a width of two Si dimers [see Fig. 1(a)].
The bottom Si layer of the slab was fully passivated by one or two H atoms per Si atom.
For simulations of finitely long DB wires, we used a supercell with the lattice vectors
of 4$\vec{a}$, $\vec{b}$, and $\vec{c}$, which includes eight Si atoms along the step edge. The electronic wave functions were expanded in a plane-wave basis set with a cutoff of 20 Ry, and the ${\bf k}$-space integration was done with 4${\times}$2${\times}$1 (1${\times}$2${\times}$1 for finitely long DB wires) ${\bf k}$-points in the surface Brillouin zone.~\cite{conv} All atoms except the bottom two Si layers were allowed to relax along the calculated Hellmann-Feynman forces until all residual force components were less than 1 mRy/bohr. The present calculation scheme has been successfully applied not only for the electronic and magnetic properties of various DB wires on the H-passivated C(001), Si(001), and Ge(001) surfaces~\cite{code} but also for the reaction and adsorption of various hydrocarbons on the Si(100) surface.~\cite{hydro}

\section{RESULTS}

We begin to optimize the atomic structure of the rebonded $D_B$ step of the Si(001) surface.
Figure 1(a) shows the optimized structure of the rebonded $D_B$ step, where the step edge atoms are designated as $S$.
The rebonded $D_B$ step separates upper and lower terraces containing two buckled
dimers $T_1$ and $T_2$ [see Fig. 1(a)]. We find that $S$ atoms have relatively long bonds to the step atoms (located between $S$ and $T_1$) with two different bond lengths of 2.46 and 2.55 {\AA}. Furthermore, $S$ atoms are displaced up and down alternatively with a height difference of 0.66 {\AA}, in good agreement with that (0.67 {\AA}) obtained from a previous~\cite{kra} DFT calculation.
The dimer bond length $d_{\rm dimer}$ and the dimer height difference ${\Delta}h$
of $T_1$ are found to be 2.36 and 0.71 {\AA}, respectively, significantly different
from those (2.29 and 0.49 {\AA}) of $T_2$.
The relatively smaller value of ${\Delta}h$ in $T_2$ compared with $T_1$ is possibly due to a strain effect at the step edge: i.e., half of the subsurface atoms attached
to $T_2$ retain bulk-like bonding arrangements so that $T_2$ has too much strain for
buckling. We note, however, that $d_{\rm dimer}$ and ${\Delta}h$ of $T_1$ are close to those (2.36 and 0.71 {\AA}) of buckled dimers at the flat Si(001) surface.

Each $S$ atom at the $D_B$ step has one unpaired electron, forming a 1D array of DBs
along the step edge. This kind of DB wire is similar to the case fabricated by the selective removal of H atoms from an H-passivated Si(001) surface along one side of an Si dimer row.~\cite{wat,hit} For the latter DB wires of finite lengths, recent DFT calculations~\cite{code,rob} predicted that the AFM configuration is energetically favored over the NM one and the energy difference between the two configurations varies with respect to the wire length.~\cite{foot} In order to examine the possibility of magnetic orders at the $D_B$ step, we determine the atomic structure of the $D_B$ step within the AFM configuration.
The optimized structure of the AFM configuration is shown in Fig. 1(b). Unlike the NM configuration showing a Jahn-Teller-like distortion, the AFM configuration is found to suppress the up and down displacements of S atoms, indicating
the absence of spin-Peierls instability. We find that the NM configuration is more
stable than the AFM configuration by 28 meV/DB.
This result is different from our previous~\cite{lee} result for the DB wire formed on the H-passivated Si(001) surface, where the AFM configuration is slightly favored over the NM configuration by 8 meV/DB. This different energetics of the NM and AFM configurations
between the $D_B$ step and the H-passivated Si(001) surface may be attributed to more delocalized character of DB electrons at the $D_B$ step, as discussed below.

\begin{figure}[hb]
\centering{ \includegraphics[width=7.0cm]{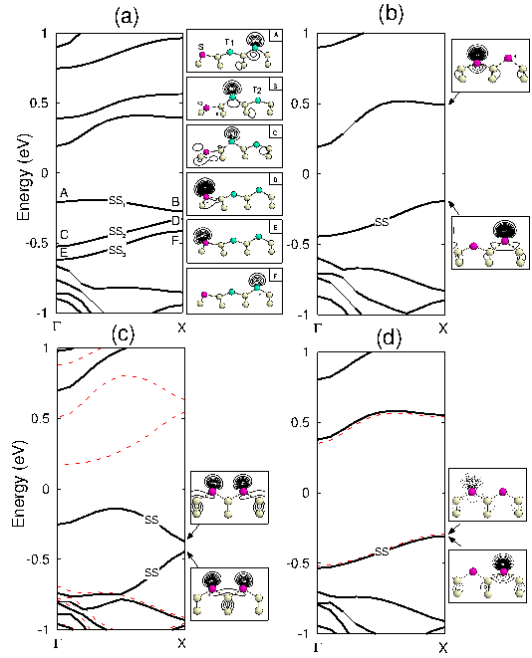} }
\caption{(color online)
 (a) Calculated band structure of the rebonded $D_B$ step, together with the charge characters of three surface states ($SS_1$, $SS_2$, and $SS_3$) at the ${\Gamma}$ and $X$ points. The first contour line and the line spacings are the same as 0.001 electron/bohr$^3$. The energy zero represents the Fermi level.
 The direction of ${\Gamma}$-$X$ line is along the step.
 The results for an infinitely long DB wire created by the H-passivation of terrace are
 also given in (b), (c), and (d) for the NM, FM, and AFM configurations, respectively. In (c) and (d), the solid (dashed) lines represent the subbands of
 up (down) spin. Charge contour plots in (b), (c), and (d) are drawn in a vertical plane
 containing S atoms.
}
\end{figure}

Figure 2(a) shows the electronic band structure of the $D_B$ step, together with the charge characters of three surface states ($SS_1$, $SS_2$, and $SS_3$) located below the Fermi level.
It is seen that the three surface states have different charge characters depending on the ${\bf k}$ vectors. For instance, the charge character of $SS_1$ at the ${\Gamma}$ (X) point represents
the DB state of $T_2$ ($T_1$), indicating a hybridization of the DB states originating from $T_1$ and $T_2$. In this way,
$SS_2$ ($SS_3$) involves a hybridization of the DB states originating from $S$ and $T_1$ ($S$ and $T_2$). Thus, we can say that DB electrons at the $D_B$ step are much delocalized to favor the NM configuration rather than the AFM configuration.

Recently, several theoretical studies predicted surface magnetism induced by
the adsorption of hydrogen atoms on surfaces.~\cite{lee,ama,oka}
Amanna $et$ $al$.~\cite{ama} reported an H-induced
magnetic order on the Pd(110) surface. Okada $et$ $al$.~\cite{oka} predicted a ferrimagnetic spin ordering in various DB networks created on an H-passivated Si(111) surface. Similarly, Lee $et$ $al$.~\cite{lee} predicted an AFM spin ordering in DB wires fabricated on an H-passivated Si(001) surface. On the basis of these previous studies, we speculate that the delocalization of DB electrons at the $D_B$ step could be prevented by passivating the terrace with atomic H [see Fig. 1(c) and 1(d)], thereby giving rise to the stabilization of magnetic order: that is, the H-passivation of terrace gives a confinement of DB electrons at the step edge, producing
magnetic order due to an enhanced electron-electron interaction. The fabrication of
such a DB wire can be realized by passivating a stepped Si(001) surface with
atomic H and then removing H atoms along the step edge via the scanning tunneling microscope nanolithography technique.~\cite{lyd,she,her} This technique has been utilized to fabricate the DB wires on the H-passivated Si(001) surface.~\cite{hit,wat}

We optimize the atomic structure of an infinitely long DB wire (designated as
DB-$\infty$) formed on the $D_B$ step within the NM, FM, and AFM configurations.
We find that the NM configuration shows alternating
up and down displacements of $S$ atoms with a height difference of 0.77 {\AA},
whereas the FM and AFM configurations have a nearly unbuckled structure.
The AFM configuration is found to be more stable than the NM (FM) configuration
by 14 (48) meV/DB. Here, the total spin of the AFM configuration is zero, while that of the FM configuration is 1 ${\mu}_{\rm B}$/DB. Figure 2(b), 2(c), and 2(d)
show the calculated band structures of the NM, FM, and AFM configurations, respectively. The charge characters of the DB state (designated as $SS$) reveal that
the NM configuration has a charge transfer from the down to the up $S$ atoms
[see Fig. 2(b)] while the AFM configuration involves an opposite spin orientation between adjacent DBs [see Fig. 2(d)].
We note that the band gap of the AFM configuration is 0.64 eV, larger than those (0.31 and 0.38 eV) of the FM and NM configurations.
This manifests that the electronic energy gain by AFM spin ordering is greater than that by FM spin ordering or Peierls-like lattice distortion.

It was reported that in the DB wire fabricated on an H-passivated Si(001) surface, electron-electron correlations are competing with electron-lattice couplings due to Peierls instability or Jahn-Teller effect.~\cite{rob,code,lee} To find the on-site electron-electron interactions in the DB wire [Fig. 1(d)] formed on the $D_B$ step, we evaluate the Hubbard correlation energy $U$ by using the constrained DFT calculations.~\cite{ani} Here, we simulate the transfer of one spin-down electron from the $SS$ state [in Fig. 2(d)] to the lowest unoccupied state and calculate the change in the total energy. We obtain $U$ = 1.71 eV,
which is much larger than the electron
hopping parameter $t$ = 0.23 eV (estimated from the bandwidth of the $SS$ state). The resulting large ratio of $U/t$ = 7.43 for an infinitely long DB wire indicates the importance
of electron correlations, giving rise to the preference of an AFM spin ordering over
a Peierls-like distortion. From the energy difference between the FM and AFM configurations,
the exchange coupling constant $J$ between adjacent spins can be estimated as 48 meV.
Note that $S$ atoms with highly localized DB electrons are covalently bonded to the subsurface Si atoms which have a negligible spin polarization.
Thus, it is most likely that an individual DB is fully spin-polarized by an intra-atomic exchange, and such magnetic moments of $S$ atoms
are antiferromangnetically coupled with each other via the superexchange mechanism,
similar to the case of the DB wires formed on the H-passivated Si(001) surface.~\cite{code,lee,rob}

\vspace{1.0cm}
\begin{figure}[ht]
\centering{ \includegraphics[width=7.0cm]{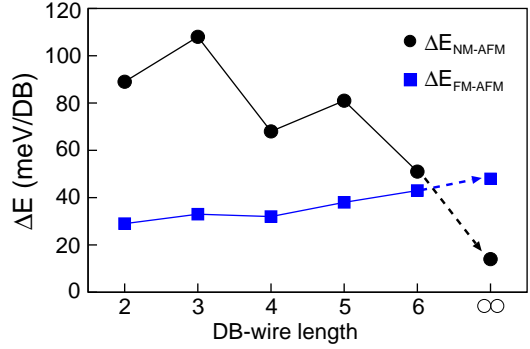} }
\caption{(color online)
Calculated total energies of the NM and FM configurations relative to the AFM
configuration for finitely long DB wires as a function of the wire length.
The results for DB-${\infty}$ are also given.
}
\end{figure}

In order to see how the magnetic stability of the DB wires varies with respect to the wire length, we determine the atomic structures of finitely long DB wires
within the NM, FM, and AFM configurations.
Here, we consider finitely long DB wires containing up to six DBs (designated as DB-2, DB-3, DB-4, DB-5, and DB-6, respectively).
The calculated total energies of the NM and FM configurations relative to the AFM configuration as a function of the wire length are plotted in Fig. 3.
We find that the AFM configuration is more stable than the NM configuration by ${\Delta}E_{\rm NM-AFM}$ = 89, 108, 68, 81, and 51 meV for DB-2, DB-3, DB-4,
DB-5, and DB-6, respectively. Thus, the energy preference of AFM ordering over Jahn-Teller-like distortion tends to increase as the wire length decreases (see Fig. 3).
This may be ascribed to the fact that the two ends in shorter DB wires
give rise to a relatively larger elastic-energy cost for the alternating up and down displacements in the NM configuratin.
It is noteworthy that ${\Delta}E_{\rm NM-AFM}$ for DB3 (DB5) is larger than those of DB2 and DB4 (DB4 and DB6). Such an even-odd oscillatory behavior correlates with the variation of DB states with respect to the wire length, as discussed below.
Figure 3 shows that the AFM configuration is energetically favored over the FM one by
29, 33, 32, 38, 43, and 48 meV/DB for DB-2, DB-3, DB-4,
DB-5, DB-6, and DB-${\infty}$, respectively. Using the Heisenberg spin Hamiltonian
$\hat{H} = J\sum_{i,j} {\bf S}_i\cdot{\bf S}_{j}$ where we assume only the existence of the nearest neighbor interactions in DB wires,
we obtain $J$ = 57, 49, 42, 47, 52, and 48 meV for DB-2, DB-3, DB-4,
DB-5, DB-6, and DB-${\infty}$, respectively.~\cite{foot2} This result indicates that the exchange
coupling constant between adjacent spins varies slightly with respect to the wire length.

\begin{figure}[ht]
\centering{ \includegraphics[width=7.0cm]{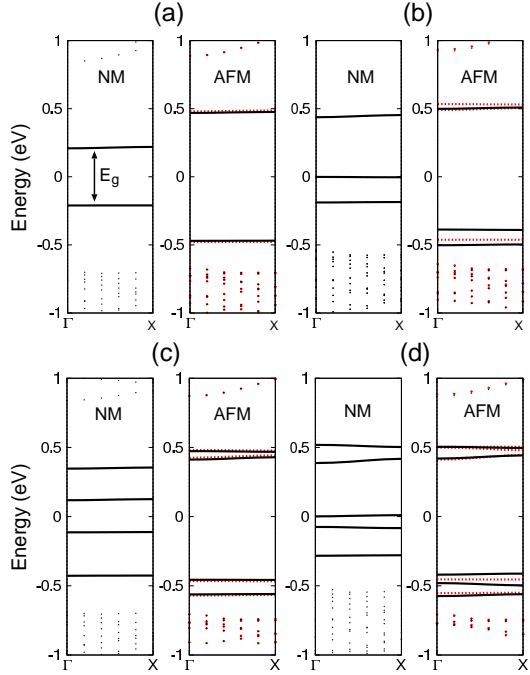} }
\caption{(color online)
Calculated band structures of (a) DB-2, (b) DB-3, (c) DB-4, and (d) DB-5 within the NM and AFM configurations. The energy zero represents the Fermi level. The direction of ${\Gamma}$-$X$ line is along the DB wire. The lines represent the subbands due to DB electrons. In the AFM configuration, the subbands of up (down) spin are drawn with the solid (dashed) lines.
}
\end{figure}

\begin{table}[ht]
 \caption{
 Calculated band gaps (in eV) for the DB-2, DB-3, DB-4, DB-5, DB-6, and DB-${\infty}$
 wires within the NM and AFM configurations.
 }
 \begin{ruledtabular}
 \begin{tabular}{ccccccc}
       & DB-2   &   DB-3  &   DB-4  &   DB-5 &   DB-6  &    DB-$\infty$  \\
 \hline
 NM    & 0.43   &   no    &   0.23  &  no    &   0.24  &     0.38        \\
 AFM    & 0.94   &  0.89   &   0.87  &  0.82  &   0.80  &    0.63        \\
 \end{tabular}
 \end{ruledtabular}
 \end{table}

Figure 4(a), 4(b), 4(c), and 4(d) show the band structures for DB-2, DB-3, DB-4, and
DB-5, respectively.
It is seen that the subbands due to DB electrons exist near $E_F$.
We find that, as the wire length increases, the subbands are lower in energy and their spacings decrease, showing the evolution of quantum well states. For DB-3 and DB-5, the NM configuration has a half-filled subband, while the AFM configuration has a band-gap opening of 0.89 and 0.82 eV, respectively (see Table I).
On the other hand, for the even-numbered DB wires of DB-2, DB-4, and DB-6,
the NM configuration has a band gap of 0.43, 0.23, and 0.24 eV, respectively, which are smaller than the corresponding ones (0.94, 0.87, and 0.80 eV) of the AFM configuration.
These different features of subbands between odd- and even-numbered DB wires
represent the so-called quantum-size effects, accounting for the above-mentioned
oscillatory behavior of ${\Delta}E_{\rm NM-AF}$ with respect to the wire length (see Fig. 3). The calculated band gaps for finitely long DB wires
also manifest that the electronic energy gain by the AFM spin polarization is larger than that by the Jahn-Teller-like lattice distortion.

\section{SUMMARY}

Our spin-polarized DFT calculations predicted that the rebonded $D_B$ step of the Si(001) surface can stabilize an AFM order by passivating the terrace with H atoms.
We found that the H-passivation of terrace enhances the localization of DB electrons
at the step edge, leading to an AFM coupling between adjacent spins of DB electrons.
We also found that finitely long DB wires exhibit the evolution of quantum-well states
as a function of the wire length, giving rise to the even-odd oscillatory behaviors
in their band structures and energetics. The H-induced magnetism predicted here in the Si DB wires may offer a unique way to device the spintronics on semiconductor substrates. Finally, we anticipate that scanning tunneling microscopy can identify indirectly the predicted AFM order by measuring the height profile along the DB wire, because the AFM configuration involves no buckling of the Si atoms composing the DB wire.

\vspace{0.4cm}
\noindent{ACKNOWLEDGEMENT}

This work was supported by National Research Foundation of Korea (NRF) grant funded by the Korean Government (NRF-2011-0015754). The calculations were performed by KISTI supercomputing center through the strategic support program (KSC-2012-C3-11) for the supercomputing application research. J.H.L. acknowledges support from the TJ Park Foundation.

$^*$ Corresponding authors; chojh@hanyang.ac.kr


\begin{thebibliography}{99}
\bibitem{yaz} O. V. Yazyev, Rep. Prog. Phys. {\bf 73}, 056501 (2010).
\bibitem{esq} P. Esquinazi, D. Spemann, R. H\"ohne, A. Setzer, K.-H. Han, and T. Butz, Phys. Rev. Lett. {\bf 91}, 227201 (2003).
\bibitem{rod} A. V. Rode, E. G. Gamaly, A. G. Christy, J. G. Fitz Gerald, S. T. Hyde, R. G. Elliman, B. Luther-Davies, A. I. Veinger, J. Androulakis, and J. Giapintzakis, Phy. Rev. B {\bf 70} 054407 (2004).
\bibitem{bar} A. P. M. Barboza, M. H. D. Guimaraes, D. V. P. Massote and L. C. Campos, Adv. Mater. {\bf 23}, 3014 (2011).
\bibitem{lee2} K. W. Lee, and C. E. Lee, Phys. Rev. Lett. {\bf 106}, 166402 (2011).
\bibitem{choi} Y.-K. Choi, J.-H. Cho, B. Sanyal, and G. Bihlmayer, Phys. Rev. B {\bf 86}, 081415(R) (2012).
\bibitem{son} Y.-W. Son, M. L. Cohen, and S. G. Louie, Nature {\bf 444}, 347 (2006).
\bibitem{fuj} M. Fujita, L. Wakabayashi, K. Nakada and K. Kusakaba, J. Phys. Soc. Jpn. {\bf 65}, 1920 (1996).
\bibitem{nak} K. Nakada, M. Fujita, G. Dresselhaus and M. S. Dresselhaus, Phys. Rev. B {\bf 54}, 17954 (1996).
\bibitem{erw} S. C. Erwin, and F. J. Himpsel, Nat. Commun. {\bf 1}, 58 (2010); {\it ibid} New Journal of Physics, {\bf 14}, 103004 (2012).
\bibitem{cha} D. J. Chadi, Phys. Rev. Lett. {\bf 59}, 1691 (1987).

\bibitem{hoh} P. Hohenberg, W. Kohn, Phys. Rev. {\bf 136}, B864 (1964).
\bibitem{koh} W. Kohn, L. J. Sham, Phys. Rev. {\bf 140}, A1133 (1965).
\bibitem{per} J. P. Perdew, K. Burke, and M. Ernzerhof, Phys. Rev. Lett. {\bf 77}, 3865 (1996); {\bf 78} 1396(E) (1997).
\bibitem{tro} N. Troullier and J. L. Martins, Phys. Rev. B {\bf 43}, 1993 (1991).
\bibitem{conv} Additional calculations with a plane-wave-basis cutoff of 30 Ry and a ${\bf k}$-point sampling of 8${\times}$4${\times}$1 showed that the energy difference ${\Delta}E_{\rm NM-AFM}$ between the NM and AFM configurations for an infinitely long DB wire changes by less than 2 meV/DB.
\bibitem{code} J.-H. Lee and J.-H. Cho, Surf. Sci. {\bf 605}, L13 (2011), and references therein.
\bibitem{hydro} J.-H. Cho and L. Kleiman, J. Chem. Phys. {\bf 121}, 1557 (2004); J.-H. Choi and J.-H. Cho, J. Am. Chem. Soc. {\bf 128}, 3890 (2006); J.-H Lee and J.-H. Cho, Phys. Rev. B {\bf 76}, 125302 (2007); J.-H. Choi and J.-H. Cho, Phys. Rev. Lett. {\bf 98}, 246101 (2007).
\bibitem{kra} P. Kratzer, E. Pehlke, M. Scheffler, M. B. Raschke, and U. H\"ofer, Phys. Rev. Lett. {\bf 81}, 5596 (1998).
\bibitem{hit} T. Hitosugi, S. Heike, T. Onogi, T. Hashizume, S. Watanabe, Z.-Q. Li, K. Ohno, Y. Kawazoe, T. Hasegawa, and K. Kitazawa, Phys. Rev. Lett. {\bf 82}, 4034 (1999).
\bibitem{wat} M. A. Walsh and M. C. Hersam, Annual Review of Physical Chemistry,
{\bf 60}, 193 (2009), and references therein.
\bibitem{lee} J. Y. Lee, J.-H. Cho, Z. Zhang, Phys. Rev. B {\bf 80}, 155329 (2009).
\bibitem{rob} R. Robles, M. Kepenekian, S. Monturet, C. Joachim, and N. Lorente, J. Phys.: Condens. Matter {\bf 24}, 445004 (2012).
\bibitem{foot} There was a controversy about the ground state of an infinitely long DB wire created from the H-passivated Si(001) surface: i.e., recent DFT calculations (Ref. 23) with the SIESTA (VASP) code predicted that the NM configuration is slightly more stable than the AFM configuration by 5 (15) meV/DB, differing from our previous DFT result (Ref. 22) that the latter configuration is energetically favored over the former one by 8 meV/DB.
\bibitem{ama} P. Amann, M. Cordin, J. Redinger, S. D. Stolwijk, K. Zumbr\"agel, M. Donath, E. Bertel, and A. Menzel, Phys. Rev. B {\bf 85}, 094428 (2012).
\bibitem{oka} S. Okada, K. Shiraishi, and A. Oshiyama, Phys. Rev. Lett. {\bf 90}, 026803 (2003).
\bibitem{lyd} J. W. Lyding, T.-C. Shen, J. S. Hubacek, J. R. Tucker, and G. C. Abeln, Appl. Phys. Lett. {\bf 64}, 2010 (1994).
\bibitem{she} T.-C. Shen, C. Wang, G. C. Abeln, J. R. Tucker, J. W. Lyding, Ph. Avouris and R. E. Walkup, Science {\bf 268}, 1590 (1995).
\bibitem{her} M. C. Hersam, N. P. Guisinger and J. W. Lyding, J. Vac. Sci. Technol. {\bf A18}, 1349 (2000).
\bibitem{ani} V. I. Anisimov, I. V. Solovyev, M. A. Korotin, M. T. Czy\`zyk and G. A. Sawatzky, Phys. Rev. B {\bf 48}, 16929 (1993).
\bibitem{foot2} The exchange coupling constant in the DB-$n$ wire is evaluated by using $J$ = $E_{\rm FM-AFM}/(n-1)$, where $E_{\rm FM-AFM}$ is the energy difference per unit cell between the FM and AFM configurations. For DB-${\infty}$, the value of $n$ is equal to 2.
\end{thebibliography}
\end{document}